\documentclass{pasj01}
\Received{$\langle$reception date$\rangle$}
\Accepted{$\langle$acception date$\rangle$}
\Published{$\langle$publication date$\rangle$}
\newcommand{{\maxi}}{{\it MAXI}}
\newcommand{{\swift}}{{\it Swift}}

\newcommand{\Tin}{T_\mathrm{in}}

\newcommand{\Rin}{R_\mathrm{in}}

\newcommand{\msolar}{M_{\odot}}

\usepackage{url}
\usepackage{lineno}
\usepackage{color}

\begin{document}

\title{Multiwavelength Observations of the Black Hole X-ray Binary MAXI J1820$+$070 in 
the Rebrightening Phase
}
\author{Tomohiro \textsc{Yoshitake}\altaffilmark{1}}%
\altaffiltext{1}{Department of Astronomy, Kyoto University, Kitashirakawa-Oiwake-cho, Sakyo-ku, Kyoto, Kyoto 606-8502, Japan}
\email{yoshitake@kusastro.kyoto-u.ac.jp}
\author{Megumi \textsc{Shidatsu}\altaffilmark{2}}
\altaffiltext{2}{Department of Physics, Ehime University, 
2-5, Bunkyocho, Matsuyama, Ehime 790-8577, Japan}
\author{Yoshihiro \textsc{Ueda}\altaffilmark{1}}

\author{Shin \textsc{Mineshige}\altaffilmark{1}}
\author{Katsuhiro L. \textsc{Murata}\altaffilmark{3}}
\altaffiltext{3}{Department of Physics, Tokyo Institute of Technology, 2-12-1 Ookayama, Meguro-ku, Tokyo 152-8551, Japan}
\author{Ryo \textsc{Adachi}\altaffilmark{3}}
\author{Hiroyuki \textsc{Maehara}\altaffilmark{4,5}}
\author{Daisaku \textsc{Nogami}\altaffilmark{1}}
\author{Hitoshi \textsc{Negoro}\altaffilmark{6}}
\author{Nobuyuki \textsc{Kawai}\altaffilmark{3}}
\author{Masafumi \textsc{Niwano}\altaffilmark{3}}
\author{Ryohei \textsc{Hosokawa}\altaffilmark{3}}

\altaffiltext{4}{Okayama Observatory, Kyoto University, 3037-5 Honjo, Kamogatacho, Asakuchi, Okayama 719-0232, Japan}
\altaffiltext{5}{Subaru Telescope Okayama Branch Office, National Astronomical Observatory of Japan, National Institutes of Natural Sciences, 3037-5 Honjo, Kamogata, Asakuchi, Okayama 719-0232, Japan}
\altaffiltext{6}{Department of Physics, Nihon University, 1-8-14 Kanda-Surugadai, Chiyoda-ku, Tokyo 101-8308, Japan}
\author{Tomoki \textsc{Saito}\altaffilmark{7}}
\altaffiltext{7}{Nishi-Harima Astronomical Observatory, Center for Astronomy, University of Hyogo, 407-2 Nishigaichi, Sayo-cho, Sayo, Hyogo 679-5313, Japan}
\author{Yumiko \textsc{Oasa}\altaffilmark{8,9}}
\altaffiltext{8}{Faculty of education, Saitama University, Simo-okubo 255, Sakura-ku, Saitama 338-8570, Japan}
\altaffiltext{9}{Graduate School of Science and Engineering, Saitama University, Simo-okubo 255, Sakura-ku, Saitama 338-8570, Japan}
\author{Takuya \textsc{Takarada}\altaffilmark{9,10}}
\altaffiltext{10}{Astrobiology Center, NINS, 2-21-1 Osawa, Mitaka, Tokyo 181-8588, Japan}
\author{Takumi \textsc{Shigeyoshi}\altaffilmark{8}}
\author{OISTER Collaboration}


\maketitle

\begin{abstract}

We report the results of quasi-simultaneous multiwavelength (near-infrared,
optical, UV, and X-ray) observations of the Galactic X-ray black hole
binary MAXI J1820+070 performed in 2019 May 10--13, $\sim 60$ days after the
onset of the first rebrightening phase. It showed a much larger %TY
optical-to-X-ray luminosity ratio ($\sim 8$) than in the initial
outburst epoch. The primary components of the spectral energy
distribution (SED) can be best interpreted by radiatively
inefficient accretion flow (RIAF) spectrum showing a luminosity peak
in the optical band. By comparison with theoretical calculations, we
estimate the mass accretion rate to be 
$\dot{M}/(8 L_{\rm Edd}/c^2) \sim 10^{-3}$, 
where $c$ is the light speed and $L_{\rm Edd}$ is the
Eddington luminosity. In addition to the RIAF emission, a blue
power-law component is detected in the optical-UV SED, which is most
likely synchrotron radiation from the jet. The optical spectrum taken
at the Seimei telescope shows a weak and narrow H$\alpha$ emission
line, whose emitting region is constrained to be $\gtrsim 2 \times
10^{4}$ times the gravitational radius. 
We suggest that the entire disk structure cannot be
described by a single RIAF solution but 
cooler material responsible for the H$\alpha$ emission 
must exist at the outermost region.

\end{abstract}
%% for PASJ, Scholar One
%\linenumbers
\section{Introduction} \label{sec:intro}

A Galactic black hole X-ray binary (BHXB), consisting of a stellar-mass 
black hole and a companion star, is an ideal object to study accretion 
physics. Most of the known BHXBs are transient sources; 
While usually in a faint, dormant state, they can suddenly brighten by several orders of magnitudes during recurring outburst periods.
In their outbursts they show several different states with distinct 
spectral and timing properties (for a review, see e.g., \cite{don07}, \cite{tet16}). 
At low luminosities, they stay in the low/hard state (LHS), which is characterized by a strong short-term variation typically on timescales of $\sim$0.1--10 s and a hard, power-law shaped spectrum often with an 
exponential cutoff at $\sim 100$ keV.
When the luminosity increases, they make a transition to the so-called 
high/soft state (HSS), where the spectrum is well represented by 
the Multi-color Disk (MCD) model \citep{mtd84}, 
an approximated spectral model of the standard accretion disk 
\citep{sha73}, and the short-term variation is strongly suppressed. 

An important issue in understanding black-hole accretion flow is how the standard disk 
forms and evolves with the mass accretion rate. 
The standard disk is believed to extend down to the innermost stable 
circular orbit (ISCO) in the HSS, because the observed $r_{\rm in}$ 
value of a BHXB is found to be constant over a wide range of the disk 
luminosity $L_{\rm disk}$ (e.g., \cite{ebs93}). 
By contrast, many studies suggest that the accretion disk in the LHS 
is truncated before reaching the ISCO (e.g., \cite{mks08, tms09, 
sdt11, sdt13}) and the inner part is replaced by low-density hot 
flow that is responsible for the Comptonization.
These results basically agree with a prediction from a standard 
theoretical model of black hole accretion flows (e.g., \cite{esi97}): 
when the mass accretion rate becomes sufficiently small (typically 
smaller than $\sim$0.1 times the Eddington rate), the standard disk 
is truncated and its inner region becomes radiatively inefficient 
accretion flow (RIAF). 
The model also predicts that the truncation radius increases as the
mass accretion rate decreases. 
However, detailed observational studies of the disk structure in the LHS 
have been limited to relatively high X-ray luminosities above $\sim 10^{36}$ 
erg s$^{-1}$ and hence at which luminosity the standard disk is generated is unknown. 
To answer this question, it is important to observe BHXBs in the very dim periods. 

MAXI J1820+070 (hereafter MAXI J1820) is a Galactic BHXB discovered 
with Monitor of All-sky X-ray Image (MAXI; \cite{mat09}) on 2018 
March 11 \citep{kwm18}. Soon after the discovery, the X-ray source was identified with the optical variable source ASSASN-2018ey \citep{tuk18, den18}, thanks to the refined position by {\it Swift} follow-up observations. The X-ray flux exceeded 1 Crab 
in the 2--20 keV band at the peak \citep{sdt18}. 
The high X-ray flux during the outburst, the small distance (3 kpc; \cite{gan19}), and low Galactic absorption (with a hydrogen column density of $N_{\rm H}\sim10^{21}$ cm$^{-2}$) of the source enabled extensive observations in various wavelengths (\cite{sdt18}, \cite{tuk18} for discovery and monitoring early outburst periods; e.g., 
\cite{bui19}, \cite{bei21} for X-ray spectral studies; e.g., \cite{pai19}, \cite{ma21} for short-term variability; e.g., \cite{mun20}, \cite{tor19} for optical or near-infrared spectroscopy; \cite{vel19}, \cite{pou22} for polarimetry, e.g., \cite{bri20}, \cite{atr20} for radio observations, e.g., \cite{sdt19}, \cite{tet2021}, \cite{sha21} for multi-wavelength studies). 

After the initial outburst that lasted for $\sim$200 days, MAXI J1820
has shown several rebrightenings in the optical and X-ray bands. 
In these phases, the optical-to-X-ray luminosity ratio was 
found to be very large, however, compared with that in the initial 
outburst epoch. 
This suggests that the accretion disk structure may be very different 
from the normal LHS or HSS observed in the 2018 outburst. 
In this article, we investigate the properties of the accretion 
flow in MAXI J1820 on the basis of our
quasi-simultaneous multi-wavelength observations performed around 2019 May 11, which corresponds to the decay phase in the first rebrightening.
We assume a black hole mass of $M = $ 7--8 $M_\odot$ and an inclination angle of $i = 69^\circ$--77$^\circ$, respectively, which are estimated by \citet{tor19}, 
and $D = 3$ kpc \citep{gan19}.

\section{Observations and Data Reduction}
\label{sec:obs}

We observed MAXI J1820 with Swift/XRT (X-ray), Swift/UVOT (UV), and OISTER (near-infrared to optical) on 2019 May 10--13. Optical spectroscopic observations were performed with KOOLS-IFU on Seimei telescope on 2019 May 11. Table~\ref{tab:observation} summarizes the
observation log.

\begin{table*}
  \centering
    \tbl{Observation Log}{
    \begin{tabular}{lccc}
  \hline \hline
  {\bf Observatory/Instrument (Filters))} & {\bf Date} & {\bf Exposure (ks)}& {\bf Observation ID} \\ \hline
  {\it Swift}/XRT & 2019 May 10--13 & 4 & 10627169/171/172/173 \\
  {\it Swift}/UVOT (UVM2, U, UVW1, UVW2) & 2019 May 11 & 0.1 each &10627171 \\
  OISTER (grizJHK) & 2019 May 12 & &\\
  {\it Seimei}/KOOLS-IFU & 2019 May 11 & 0.6 ($\times$15 frames)&\\
  \hline
  \end{tabular}}
  \label{tab:observation}
  \end{table*}

\subsection{\swift}

The time-averaged {\it Swift}/XRT spectrum was obtained via 
the XRT on-demand web interface\footnote{\url{https://www.swift.ac.uk/user_objects/}}.
Here, we combined all the XRT data taken from 2019 May 10--13 (OBSIDs$=$00010627169, 00010627171, 00010627172, and 00010627173) 
to improve statistics.
The {\it Swift}/UVOT data taken on 2019 May 11 were analyzed by 
using HEAsoft version 6.26.1 and the latest {\it Swift}/UVOT 
Calibration Database (CALDB) as of 2018 April. We started with 
the cleaned sky-coordinate images of the individual filters: 
UVW1, UVW2, UVM2, and U bands. We performed aperture photometry 
using the UVOT-specific tool {\tt uvot2pha} included in HEAsoft, 
where we defined the source region as a circle with a $5"$ radius 
centered at the source position and the background region as 
a circle of a $20"$ radius in a blank-sky area. 

\subsection{{\it Seimei} Telescope}
The 3.8-m Seimei telescope of Kyoto University is a new telescope 
at the Okayama Observatory \citep{krt10}. The Kyoto Okayama Optical
Low-dispersion Spectrograph with an integral field unit
(KOOLS-IFU; \citep{ysd05,mtb19}) is installed there through the optical
fibers. The data were taken under the program 19A-K-0009.  We used the
VPH-blue grism, whose wavelength coverage is 4000--8900 \AA and
wavelength resolution is $R=\lambda/\Delta \lambda \sim$ 500.

We adopted the standard data reduction procedures for optical spectra:
overscan and bias subtraction, flat fielding, wavelength calibration,
spectral extraction, sky subtraction, and flux calibration, utilizing
the Hydra package in IRAF \citep{bar94, bar95} and a python script
specifically developed for KOOLS-IFU data
reduction\footnote{\url{http://www.kusastro.kyoto-u.ac.jp/~kazuya/p-kools/reduction-201806/install_software.html}}.
We used arc lamp (Hg and Ne) data for the wavelength calibration.  The
sky subtraction was performed by using the object frames themselves.
The sky brightness was estimated with the fibers placed on blank-sky
regions. These processes were carried out for the individual object
frames. We made the final spectrum through median combination of the
spectra made from the individual object frames.
As the flux error in each spectral bin, we adopted the standard deviation of the fluxes in the line-free region around the H$\alpha$ line (6300--6500 \AA~and 6600--6800 \AA), which was estimated to be $1.4 \times 10^{-17}$ erg cm$^{-2}$ s. 

Analysis of the Seimei spectrum was carried out with XSPEC. 
Conversion of the spectral data to the XSPEC format was 
performed with the ftool {\tt ftflx2xsp}. 
To consider the spectral resolution of the {\it Seimei}/KOOLS-IFU, 
we created, using the ftool {\tt ftgenrsp}, a response matrix file 
such that the response is expressed as a Gaussian for each 
wavelength bin. A full width at half maximum (FWHM) of 9.55 \AA~was adopted for the Gaussian components, which was estimated 
from the profile of a Ne line around the H$\alpha$ line 
in the arc lamp frame. 

\subsection{OISTER}

The Optical and Infrared Synergetic Telescopes for Education and
Research (OISTER) is a framework of multi-band imaging observations
based on a Target-of-Opportunity (ToO) program. Many small-to-medium
size telescopes operated by observatories and Universities in Japan
participate in the OISTER collaboration, including the Multicolor
Imaging Telescopes for Survey and Monstrous Explosions
(MITSuME) at the Akeno Observatory \citep{kot05, mitsumeAkeno2007, mitsumeAkeno2008}
and at the Okayama Observatory \citep{mitsumeOkayama2010}, 
55 cm SaCRA telescope at Saitama University \citep{sacraMusashi2020}, and the 2.0 m 
Nayuta telescope at the Nishi-Harima Astronomical Observatory \citep{nic2011, nic2013}.

The OISTER optical and near-infrared data were reduced on IRAF by following
standard procedures including bias and dark subtraction, flat
fielding, and bad pixel masking. IRAF was also used for photometry.
The magnitudes of MAXI J1820 were calibrated with nearby reference
stars, whose magnitudes were taken from the UCAC4 catalog
\citep{2013AJ....145...44Z} and Pan-STARRS1 Surveys
\citep{2016arXiv161205560C} for the optical data, and the Two Micron
All Sky Survey Point Source Catalog \citep{2003tmc..book.....C} for
the near-IR data. Their typical errors were estimated as $\sim$10\%
of the magnitudes.

\begin{figure*}[htbp]
    \begin{center}
      \includegraphics[width=150mm]{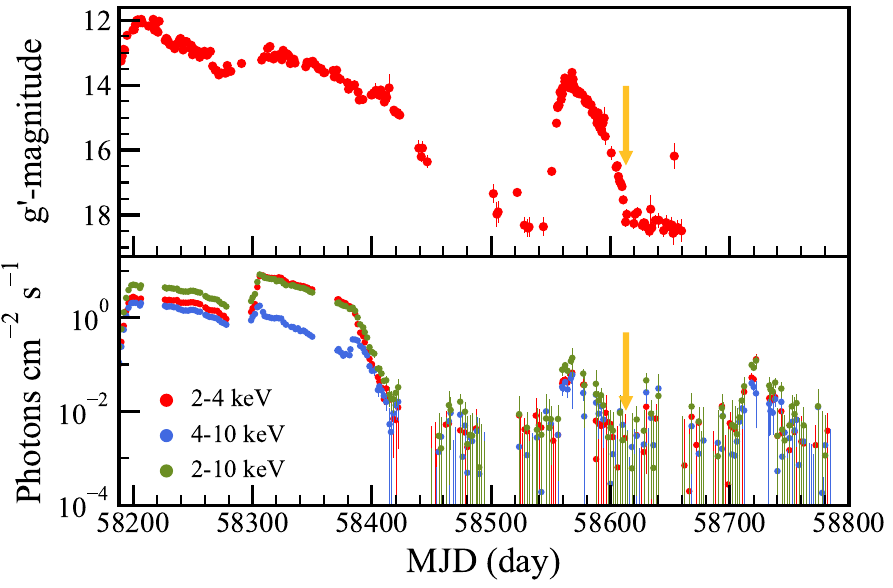} 
      \end{center}
    \caption{(Top) g'-band optical light curve of MAXI J1820 from the OISTER collaboration (Adachi et al. in prep).
    (Bottom) X-ray light curve in 2--10 keV from the MAXI/GSC. 
MJD 58200 corresponds to 2019 March 23. The yellow line presents our observation epoch (2019 May 11 $=$ MJD 58614). }
    \label{fig:light_curve}
\end{figure*}

\section{Analysis and Results}
\label{sec:ana}

\subsection{Long-term Light Curves}
\label{subsec:longterm_lc}
Fig.~\ref{fig:light_curve} shows the long-term X-ray 
and optical light curve over three years since its discovery. 
The X-ray light curve was made from the MAXI/Gas Slit Camera 
(GSC; \cite{mih11}) data via the MAXI on-demand web interface
\citep{nak13}\footnote{\url{http://maxi.riken.jp/mxondem}}. 
The optical data were taken by the OISTER collaboration 
(Adachi et al.\ in prep). The X-ray (2--10 keV)
peak flux in rebrightenings were $\sim$ 1 order of magnitude 
lower than that in the 2018 outburst, whereas the optical 
peak fluxes in these activities were comparable. Our multi-wavelength 
observation was performed in the last part of the decay phase of 
the first rebrightening, with the g'-band magnitude of $\sim 16$ mag.

\subsection{Multi-wavelength SED}

Figure~\ref{fig:unfolded_spec} shows the multi-wavelength data from the X-ray to near-infrared bands. 
We analyzed the data on XSPEC version 12.11.1. 
In the analysis, we adopted $\chi^2$ statistics for the 
infrared to UV data, whereas C-statistics for the X-ray data 
because of the low data statistics. The errors 
represent the 90\% confidence range for one parameter, 
unless otherwise stated.
We adopted the solar abundance 
table given by \citet{2000ApJ...542..914W}.

\subsubsection{X-ray Spectrum}
\label{subsubsec:ana_xspec}

We first focused on the X-ray spectrum. 
The origin of the X-ray emission 
from BHXBs at such dim phases is still unclear, although 
several possible scenarios are proposed, including 
the Bremsstrahlung radiation and/or Comptonization 
in the RIAF (e.g., \cite{man97}), or synchrotron or 
Comptonization in the jet (e.g., \cite{mar01}). 
Here, we adopted an empirical model: the simple power-law model, 
often used to characterize the LHS spectra. To account for 
the interstellar absorption, the {\tt TBabs} model was 
multiplied with $N_\mathrm{H}$ fixed at 
$1.1 \times 10^{21}$ cm$^{-2}$ \citep{2018ATel11423....1U}. 

\begin{figure*}[htbp]
  \begin{center}
     \includegraphics[width=80mm]{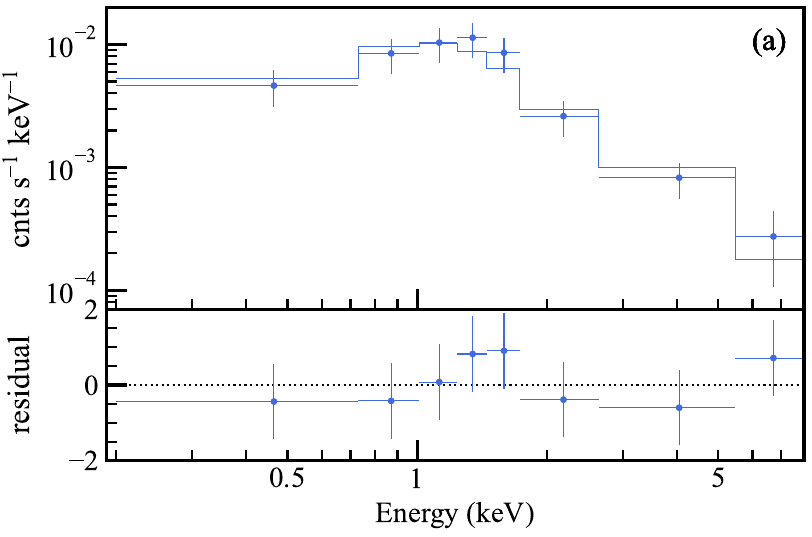} 
     \includegraphics[width=80mm]{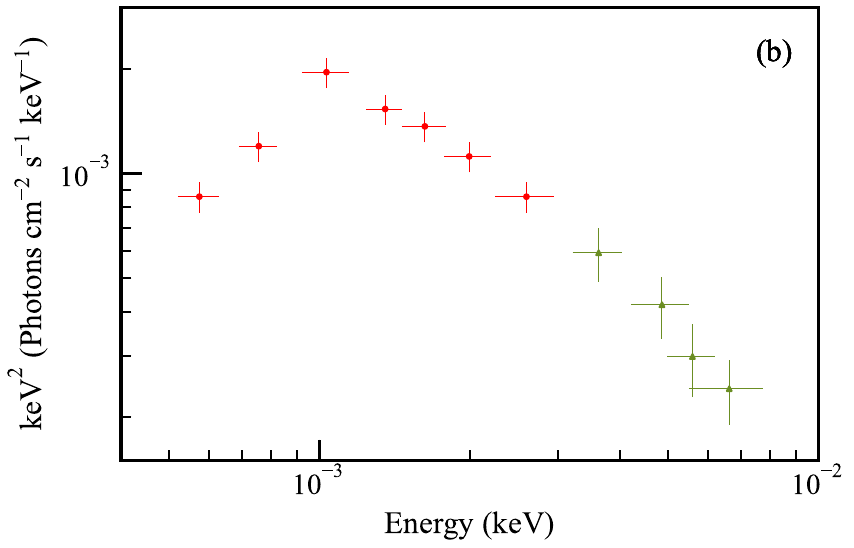} 
  \end{center}
  \caption{(a) folded (instrumental response included) X-ray spectrum with the best-fit absorbed power-law model (top) and the residuals (bottom). (b) infrared to UV SED where extinction is not corrected.}
  \label{fig:unfolded_spec}
\end{figure*}

The model successfully reproduced the X-ray spectrum. 
Figure~\ref{fig:unfolded_spec} plots the data 
and the best-fit model spectrum, and Table~\ref{tab:SEDpars} 
gives the best-fit parameters. 

\subsubsection{Infrared to UV SED}
\label{subsubsec:ana_sed}

Next, we analyzed the SED in the near-infrared (NIR), optical, and UV bands. The photons in these wavelengths from BHXBs can 
be produced in the outer region of the accretion disk,  
in the jets, and on the surface of the companion star. 
We considered two possibilities as the primary origin 
of the SED: 
(1) RIAF  
and
(2) truncated hot standard disk
(hereafter we call Model 1 and 2, respectively) and fit the SED 
with models based on these two cases in the following sections. 

In both cases, to consider the blackbody emission from the 
companion star, we added {\tt bbodyrad}, whose parameters are 
the temperature $T_{\rm bb}$ and the normalisation, 
which is determined by the radius of the emission region. 
The companion star of MAXI J1820 is suggested 
to be a K4V star \citep{2019ApJ...882L..21T}, whose surface 
temperature is $\sim 4700$ K. Hence, we fixed $T_{\rm bb}$ 
at 4700 K, and assumed a typical stellar radius of 
0.65 $R_\odot$ for a K4V star in this analysis. Note that this 
component turned out not to be significant in either case 
(see below), but since it should certainly exist,  
we included it to describe the actual physical condition.
The interstellar extinction was also taken into account 
by incorporating the {\tt redden} model 
with $E(B-V)$ at $0.16$ \citep{2018ATel11418....1B}. 
Thus, Model 1 and 2 were expressed as 
{\tt redden*(cutoffpl+bbodyrad)} 
{\tt redden*(diskbb+bbodyrad)}, respectively.

\subsubsection{Model 1: RIAF}
Theoretical studies suggests that the synchro-cyclotron emission 
of thermal electrons in the RIAF is dominant and the spectrum 
peaks around the NIR to UV band \citep{man97}. 
As an approximated model for the synchro-cyclotron 
spectrum, we adopted the cutoff power-law model {\tt cutoffpl}, 
whose parameters are the photon index 
$\Gamma$ and the cutoff energy $E_{\rm cut}$. The former 
parameter was fixed at $-0.232$, which was estimated from 
Figure 8 in \citet{man97}.

\begin{figure*}[htbp]
  \begin{center}
     \includegraphics[width=150mm]{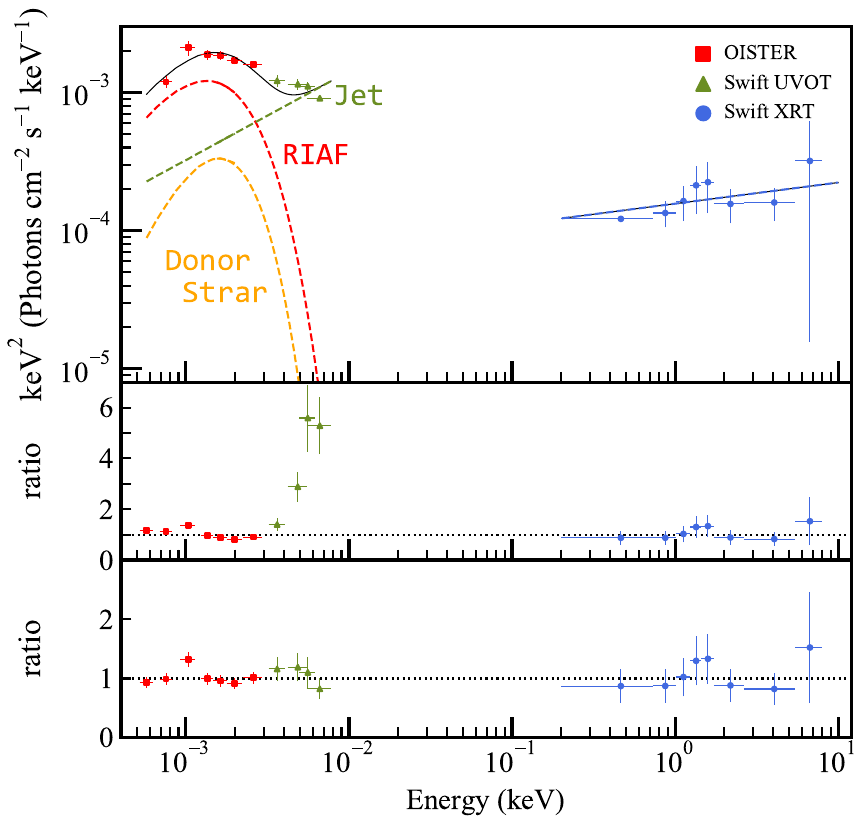} 
\end{center}
  \caption{Top: the SED data with the best-fit {\tt redden*(powerlaw+cutoffpl+bbodyrad)} model (Model 1) in the NIR to UV 
  band and the unfolded spectrum with the best-fit {\tt TBabs*power-law} model (Section~\ref{subsubsec:ana_xspec}) in X-rays, all corrected for the interstellar absorption and extinction. The individual components are separately plotted. Middle: the data versus model ratio of the {\tt redden*(cutoffpl+bbodyrad)} model (in the infrared to UV band) and the {\tt TBabs*power-law} model (in the X-ray band). Bottom: Same as the middle panel, but with the {\tt redden*(powerlaw+cutoffpl+bbodyrad)} model in the infrared to UV band.
  }
  \label{fig:best_fitting_model_RIAF}
\end{figure*}

This model, however, underestimated the observed flux 
and left large residuals in the UV band, as shown in Figure~\ref{fig:best_fitting_model_RIAF}. They gave a large 
$\chi^2$ value, 59.2, with the degree of freedom (d.o.f.) of 9. 
This suggests that additional emission components, such as 
the jet synchrotron emission and Comptonization 
components in the RIAF, are required. To account for 
these components, a power-law component was added 
to both models. The final SED model was then expressed as 
{\tt redden*(cutoffpl+powerlaw+bbodyrad)}. 
This model was found to improve the fit significantly, 
yielding $\chi^2 = 9.9$ with the d.o.f. reduced by 2.

The model successfully reproduce the actual SED profile. 
The best-fit model and its parameters are shown in 
Figure~\ref{fig:best_fitting_model_RIAF} and 
Table~\ref{tab:SEDpars}, respectively. 
The best-fit {\tt cutoffpl} component indicates 
that the synchro-cyclotron spectrum from RIAF has a peak 
at $E_{\rm cut} = 6.0^{+1.0}_{-0.6} \times 10^{-4}$ keV.

\begin{figure*}[htbp]
  \begin{center}
     \includegraphics[width=150mm]{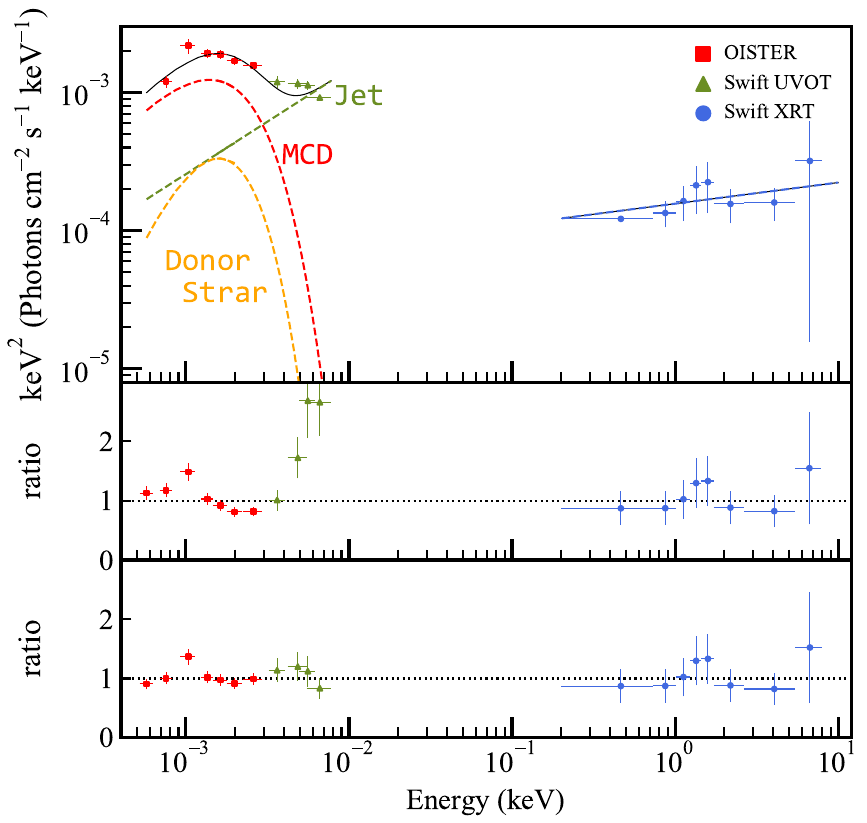}
  \end{center}
  \caption{Same as Fig.~\ref{fig:best_fitting_model_RIAF}, 
  but for the {\tt redden*(diskbb+bbodyrad)} model in the middle panel and for the {\tt redden*(powerlaw+diskbb+bbodyrad)} model (Model 2) in the top and bottom panels, in the NIR to UV band.}
  \label{fig:best_fitting_model_MCD}
\end{figure*}

\begin{table*}
    \centering
    \tbl{Best-fit parameters of the X-ray spectrum and the NIR to UV SED}{
    \begin{tabular}{lccccc}
    \hline \hline
    Component & Parameter & Unit & \multicolumn{3}{c}{Value}  \\
    & & & X-ray\footnotemark[$*$] & NIR--UV (Model 1)\footnotemark[$\dagger$] & NIR--UV (Model 2)\footnotemark[$\dagger$] \\ \hline
    {\tt powerlaw} 
        & $\Gamma$ & & $1.8 \pm 0.3$ & $ 1.4^{+0.4}_{-0.9} $ & $1.2^{+0.5}_{-1.2}$\\
    & Flux  &$10^{-12}$ ergs cm$^{-2}$ s$^{-1}$  & $6.3^{+1.4}_{-1.3} \times 10^{-1} $  & $1.8^{+0.9}_{-1.0}$ & $1.5 \pm 1.0$\\
    & $L_{\rm pow}^{\rm X,opt}$\footnotemark[$\ddagger$] & erg s$^{-1}$ & $6 \times 10^{32}$ & $2 \times 10^{33}$ & $2 \times 10^{33}$ \\
    {\tt cutoffpl} & $\Gamma$ &  & - & $-0.232$ (fixed) &- \\
    & $E_{\rm cut}$ & keV & - & $6.0_{-0.6}^{+1.0}\times 10^{-4}$ & - \\   
     & norm  & & - & $3.0_{-1.5}^{+1.1} \times 10^4$  & - \\  
    & $L_{\rm cpl}$\footnotemark[$\ddagger$] & erg s$^{-1}$ & - & $3 \times 10^{33}$ & - \\
    {\tt diskbb} & $T_{\rm{in}}$ & keV & - & - & $5.9^{+1.0}_{-0.7} \times 10^{-4}$\\
    & $R_\mathrm{in,SED}$\footnotemark[$\S$]  & km & - & - & $6.3_{-2.3}^{+1.7}\times 10^5$ \\ 
    & $L_{\rm dbb}$\footnotemark[$\ddagger$] & erg s$^{-1}$ & - & - & $3 \times 10^{33}$ \\ \hline
    C-stat/d.o.f., $\chi^2/{\rm d.o.f.}$\footnotemark[$\ddagger$] & & & 60/154 & 9.9/7 & 11.4/7 \\ 
    \hline
    \hline
    \end{tabular}}
    \begin{tabnote}
    \hangindent6pt\noindent
      \hbox to6pt{\footnotemark[$*$]\hss}\unskip% 
            {\tt TBabs*powerlaw}, with $N_\mathrm{H}$ of {\tt TBabs} fixed at $1.1 \times 10^{21}$ cm$^{-2}$. \\
    \hangindent6pt\noindent
    \hbox to6pt{\footnotemark[$\dagger$]\hss}\unskip% 
      {\tt redden*(cutoffpl+powerlaw+bbodyrad)} (Model 1) and 
      {\tt redden*(diskbb+powerlaw+bbodyrad)}(Model 2), with $E(B-V)$ of {\tt redden} fixed at $0.16$. For the {\tt bbodyrad} component, $T_{\rm bb} = 4700$ K and $R_{\rm bb} = 0.65 R_\odot$, and $D = 3$ kpc were assumed. \par
    \hangindent6pt\noindent
     \hbox to6pt{\footnotemark[$\ddagger$]\hss}\unskip% 
      Absorption/extinction corrected luminosity of each component in 0.5--5 keV (for the X-ray data) and 0.5--5 eV (for the NIR--UV data). In calculating the luminosity of the {\tt diskbb} component, $D = 3$ kpc and $i = 70^\circ$ are assumed. 
    \hangindent6pt\noindent
     \hbox to6pt{\footnotemark[$\S$]\hss}\unskip% 
      $D = 3$ kpc, $i = 70^\circ$, and $M = 7 M_\odot$ are assumed. 
      \\
    \hangindent6pt\noindent
     \hbox to6pt{\footnotemark[$\#$]\hss}\unskip% 
      Cash statistics was used for the X-ray spectrum, whereas $\chi^2$ statistics for the NIR to UV SED. Here, C-stat/d.o.f. and $\chi^2$/d.o.f. are shown for the results from the former and latter data, respectively.    
      \\
    \end{tabnote}
    \label{tab:SEDpars}
  \end{table*}

\subsubsection{Model 2: Truncated Hot Standard Disk}
In Model 2, we assumed the standard disk emission 
dominates the NIR to UV flux, for which we 
used the MCD model (the {\tt diskbb} model on 
XSPEC \citep{mtd84}). Its parameters are the 
inner disk temperature $T_{\rm{in}}$ and the 
normalization, which is related to the inner disk radius $R_{\rm{in}}$. 
As in Model 1, this model underestimated the UV flux (Fig.~\ref{fig:best_fitting_model_MCD}), 
yielding a large $\chi^2$ value of 45.7, with d.o.f. $=$ 9. 
We then combined an additional power-law component  
and adopted the model {\tt redden*(diskbb+powerlaw+bbodyrad)}.

This model improved the fit and reproduced the SED 
equally well compared with Model 1, giving $\chi^2 = 11.4$ 
with a decrease in d.o.f. by 2. The best-fit model is 
shown in Figure~\ref{fig:best_fitting_model_MCD} and 
the resultant parameters are listed in 
Table~\ref{tab:SEDpars}.
We obtained $T_{\rm{in}}  \approx 5.9 \times 
10^{-4}$ keV (i.e., $\approx 6.8 \times 10^3$ K) 
 and $R_{\rm{in}} \approx 6.3 \times 10^5$ km. 
 This suggests that the standard disk was highly truncated 
 and replaced by the RIAF inside it. 
 
BHXBs often show significant irradiation of X-rays on the outer disk 
region in their outburst periods. To investigate the effects by the X-ray illumination in the infrared to UV band, we tested an irradiated disk model. 
Here, we only considered the illumination by the inner disk emission and ignored that by the jet emission, which would cause only a limited effect due to the relativistic beaming. According to \citet{mil06}, jets in BHXBs typically have a bulk Lorentz factor of $\Gamma_\mathrm{jet} \gtrsim 10$. Considering the luminosity is proportional to $\delta^4$, where $\delta$ is the beaming factor expressed with the angle $\theta$ from the jet axis as $\delta = \Gamma_\mathrm{jet}^{-1}(1-\beta \cos\theta)^{-1}$, we found that the flux of the jet emission illuminating the outer disk (at $\theta \gtrsim 90^{\circ}$) was smaller by a factor of $\gtrsim 5$ than that for the observers at $\theta \sim 70^{\circ}$.

We employed the irradiated disk model diskir \citep{gie08,gie09} and replaced the {\tt diskbb} model in Model 2 with it to fit the SED. 
The {\tt diskir} model calculates an irradiated disk spectrum, where the X-ray photons are produced by the MCD emission and its Comptonization. Here, we ignored the illumination of the inner disk region (i.e., $f_\mathrm{in} = 0$, where $f_\mathrm{in}$ is the fraction of the luminosity of the Comptonization component that is thermalized in the inner disk), and fixed $T_\mathrm{e}$ at 100 keV, 
$f_\mathrm{out}$ at 0.01, and $\log(R_\mathrm{out}/R_\mathrm{in})$ at 0.69 (where $T_\mathrm{e}$, $f_\mathrm{out}$, and $R_\mathrm{out}$ are  
the electron temperature of the Comptonization component, the fraction of the bolometric flux which is thermalized in the outer disk, and the outer disk radius, respectively). We note that the fit results unchanged when  $f_\mathrm{out} = 0.1$ and 0.001 were adopted. For $R_\mathrm{out}$, we employed the Roche lobe size of the black hole, $3.2 \times 10^6$ km, which was calculated from Equation (4) in \citet{pat71}, while we assumed $\Rin = 6.3 \times 10^4$ km, which was estimated from the normalization of the best-fit {\tt diskbb} model. The other parameters: the inner disk temperature $\Tin$, the photon index $\Gamma$, the luminosity ratio  $L_\mathrm{c}/L_\mathrm{d}$ (where $L_\mathrm{c}$ and $L_\mathrm{d}$ represents the luminosity of the Comptonization and disk components, respectively), and the normalization, whose definition is the same 
as that of {\tt diskbb}, were allowed to vary.

The resultant parameters, however, unchanged from the best-fit {\tt diskbb} model within the 90\% confidence ranges, suggesting that X-ray irradiation makes only a negligible effect to the infrared to UV continuum. From this model, we obtained $\Tin = 5.9^{+1.2}_{-0.7} \times 10^{-4}$ keV, $\Gamma = 1.9^{+0.4}_{-0.3}$, $L_\mathrm{c}/L_\mathrm{d} = 1.1^{+1.2}_{-0.2}$, and the normalization of $(2 \pm 1) \times 10^{12}$, which corresponds to $R_\mathrm{in} = (7 \pm 2) \times 10^5$ ($D/3$ kpc) $(\cos i/\cos 70^\circ)^{-1/2}$ km.

\subsection{Optical Spectrum}

\begin{figure*}[htbp]
  \begin{center}
     \includegraphics[width=150mm]{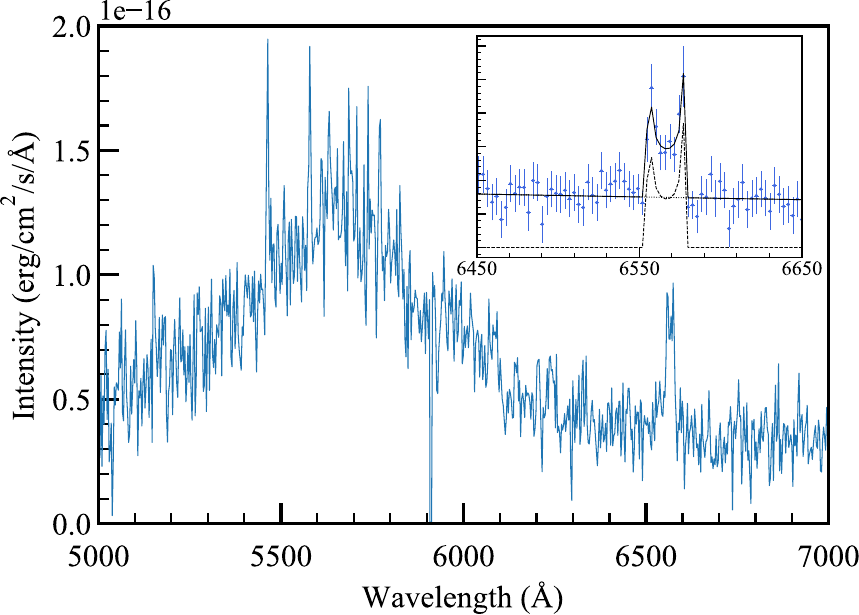} 
  \end{center}
  \caption{The Seimei/KOOLS-IFU spectrum of MAXI J1820, taken on 2019 May 11. The magnified image presents the Seimei spectrum around the $H\alpha$ line 
  and the best-fit {\tt diskline} model.
 } 
  \label{fig:optical_spectrum}
\end{figure*}

Figure \ref{fig:optical_spectrum} shows the Seimei spectrum of MAXI J1820.
The source signals are significantly detected over 5000--7000 \AA, 
in which the $\mathrm{{H\alpha}}$ emission line ($6563$ \AA)
is clearly seen. 

We performed spectral fitting on XSPEC, adopting the data around the H$\alpha$ line in a wavelength range of 6300--6700 \AA. 
First, we applied a single 
Gaussian and a power-law model for the line and the continuum, respectively. The spectral model gave an acceptable fit, yielding $\chi^2$/d.o.f $= 72/137$ and a line-center wavelength of $6565 \pm 3$ \AA and a 1-sigma line width of $11\pm 4$ \AA. 
Next, we replaced the single Gaussian with a double Gaussian at
different energies. This model is often used to approximately
represent the profile of a line emitted from an accretion disk (e.g., \cite{tet2021}). 
The best-fit model ($\chi^2$/d.o.f=67/134) gave the double-peak
separation of 14 \AA, FWHM of 20 \AA~and the half width at zero intensity (HWZI) of 19
\AA, where the zero intensity was measured at 5$\sigma$
level of a Gaussian. It is possible to roughly estimate the innermost
radius at which the line is emitted as $R_{\rm in} = 
(c \sin (i)/v_{\rm in})^2 r_{\rm g}$, where $c$ is the light velocity, $i$ the inclination, 
$v_{\rm in}$ 
the velocity at the inner edge of the disk, and $r_{\rm g} \equiv
GM/c^2$ 
($G$ is the gravitational
constant) the gravitational radius (see \cite{ber2016}). Using the HWZI value obtained from the double Gaussian model to
estimate $v_{\rm in}$, we obtained $R_{\rm in} = 1.0\times10^5 
r_{\rm g}$ for $i=70^\circ$. 

Finally, to most accurately estimate the line emitting radii directly from the data,  
we apply a physical disk emission line model, {\tt diskline}, in place of the Gaussian models.
The {\tt diskline} model computes the emission line from the  
accretion disk illuminated by the X-rays emitted in the inner disk region, with input parameters of the line energy, the inner and outer disk radii of the emission region ($R_{\rm in}$ and $R_{\rm out}$, respectively), the inclination angle ($i$), and the power law index ($\alpha$) of emissivity law ($\propto E^\alpha$). We adopted $\alpha = -3$, assuming the X-ray source is a point source located at a finite height above the black hole. We fixed $i$ and $R_{\rm out}$ at 70$^\circ$ \citep{tor20} and $3 \times 10^5 r_{\rm g}$,
respectively. The latter value corresponds to the Roche lobe size of the black hole in MAXI J1820, estimated from Kepler's third law and Equation 4 in \citet{pac71}, assuming an orbital period of 0.68 day and black hole and companion star masses of $7 M_\odot$ and $0.5 M_\odot$, respectively \citep{tor20}. 
The model gave $R_{\rm in} \gtrsim 2 \times 10^4 r_{\rm g}$
at a 90\% confidence limit,   
with the fit quality $\chi^2/$d.o.f = $68/137$ comparable to 
those
obtained with the single and double Gaussian models.
The lower limit on the innermost radius is consistent with 
the best-fit value estimated with
the double Gaussian model (see above).
The best-fit model and the data are shown in Figure~\ref{fig:optical_spectrum}. 
The equivalent width of the line is estimated to be $34\pm 6$ \AA.

\section{Discussion}
\label{sec:discussion}

\subsection{SED Modeling}

We have obtained quasi-simultaneous multiwavelength SED of MAXI  J1820 
about 60 days after the onset of the first re-brightening phase,
covering the near-IR, optical, UV and X-ray bands.
We find that the optical-to-X-ray luminosity ratio
is much larger than those in the LHS and HSS observed in the initial
outburst epoch, as summarized in Table~3. This suggests very different
physical conditions of the accretion flow in our epoch compared with
these canonical states.

\begin{table*}
  \centering
%  \small
  \tbl{Comparison with the state of MAXI J1820 in each work.}{
  \label{tab:comparison}
  \begin{tabular}{lcccl}
  \hline \hline
  State & MJD & $L_{\rm X}$\footnotemark[$*$] (erg sec$^{-1}$) & $L_{\rm opt}/L_{\rm X}$\footnotemark[$*$] & Reference \\
  HSS & 58314 & $4\times 10^{37}$ & $6 \times 10^{-3}$ &  \citet{sdt19} \\
  LHS & 58201 & $3\times 10^{37}$ & $4 \times 10^{-2}$  & \citet{sdt18} \\
  RIAF & 58614 & $7\times 10^{32}$ & $8$ & This work \\
  \hline
\end{tabular}}
    \begin{tabnote}
    \hangindent6pt\noindent
      \hbox to6pt{\footnotemark[$*$]\hss}\unskip% 
      $L_{\rm X}$ and $L_{\rm opt}$ are absorption/extinction corrected luminosity in the 0.5--5 keV and 0.5--5 eV bands, respectively. 
      \\
      \end{tabnote}
\end{table*}

The main part of the optical-to-UV SED 
can be described by either 
of the two models: 
(Model 1) synchro-cyclotron emission by thermal electrons in the RIAF, 
which is approximated by a cutoff power-law model, 
or (Model 2) emission from a truncated hot standard disk, 
represented by the MCD model. 
In the following, we show that Model 1 gives 
a more physically plausible interpretation than Model 2. 

\subsubsection{RIAF Interpretation (Model 1)}
\label{sec:discussion1}

The observed low X-ray luminosity, $L_ {\rm pow}^{\rm X} \sim 7 \times 10^{32}$ erg s$^{-1}$,
implies a very low mass accretion rate in the inner part of the disk.
Then, it is expected that the accretion disk should become RIAF, such
as Advection-Dominated Accretion Flow (ADAF; e.g., \citet{nar95}).
Theoretical calculations of ADAF spectra (e.g., \citet{mah97, man97}) show that the SED covers a wide range of frequencies
over 8 orders of magnitude, consisting of multiple components; the
radio-to-optical emission is produced by synchro-cyclotron radiation
by hot electrons in the ADAF, whereas the X-ray emission is mainly
produced by thermal Bremsstrahlung, with an additional contribution of
a Comptonized component of the synchro-cyclotron photons, by the same hot
electrons.

To compare our SED with a model prediction, we refer to the theoretical
calculation by \citet{man97}. Their Figure~8 plots a typical SED of
ADAF for a black hole mass of $m (\equiv M/\msolar) = 10$ and
a normalized mass accretion rate of $\dot{m} (\equiv
\dot{M} c^2/8 L_{\rm Edd} ) =10^{-4}$, where 
$L_{\rm Edd}$ is the Eddington luminosity. 
It shows a sharp luminosity peak in the optical band
and a power-law like spectrum with $\Gamma \simeq 1.8$ in the X-ray band,
which are similar to the observed SED. 
Considering that this ADAF model is a very simplified one (e.g., ignoring 
outflow), we do not attempt to directly fit the model to our
broadband SED, which is beyond the scope of this paper.
Instead, we investigate if the main
properties of the observed SED (peak frequency, optical and X-ray luminosities)
can be roughly reproduced by this model only by varying
the mass accretion rate.
\citet{mah97} summarized the scaling laws for ADAF that (1)
the luminosity of the synchro-cyclotron radiation $\propto m^{0.5} \dot{m}^{1.5}$,
(2) its peak frequency $\nu_{\rm P} \propto m^{-0.5} \dot{m}^{0.5}$, and
(3) the luminosity of the Bremsstrahlung component $\propto m \dot{m}^{2}$.
Thus, if we take $m=7$ and $\dot{m}=10^{-3}$, 
$L_{\rm RIAF}$ (0.5--5 eV) $\sim 1.3 \times 10^{34}$ ergs s$^{-1}$, $\log \nu_{\rm P} \sim 15$, 
and $L_{\rm RIAF}$ (0.5--5 keV) $\sim 1.2 \times 10^{32}$ ergs s$^{-1}$
are predicted.
These values are all consistent with those of the observed SED
($L_{\rm cpl} \sim 3 \times 10^{33}$ ergs s$^{-1}$, $\log \nu_{\rm P} \sim 14.2$, $L_{\rm pow}^{\rm X} \sim 7 \times 10^{32}$ ergs s$^{-1}$; see Table~2).
within factors of 6, even if we do not apply any tuning for the other
parameters (viscosity and magnetic field strength). Thus, the RIAF
interpretation is very plausible as the primary origin of the optical
and X-ray SED.

\subsubsection{Truncated Hot Standard-Disk Interpretation (Model 2)}

As an alternative scenario, a major part of the optical SED could be reproduced by
optically-thick disk emission. From the best-fit parameters of the MCD
component, we are able to estimate the innermost radius of the
optically thick disk.  For simplicity, here we ignore possible color
correction (i.e., the ratio between the color and effective
temperatures is assumed to be unity) in deriving the radius. We obtain
\begin{linenomath}$$
R_\mathrm{{in}} = 6.3 \times 10^{5} \left(\frac{\cos 70^\circ}{\cos i} \right)^{\frac{1}{2}}
\frac{D}{3 {\rm kpc}} {\rm km},
$$ \end{linenomath}
which corresponds to $1.1 \times 10^{4} r_{\rm g}$ for a black hole mass of
7 $M_\cdot$. 
This disk truncation radius is much larger
the truncation radius of MAXI J1820 in
the LHS estimated by \citet{sdt18}, 12--36$r_{\rm g}$.
We note that it is even much
larger than that observed in the BHXB XTE J1118+480 in its LHS
\citep{cha10}, which showed $R_\mathrm{{in}} \sim 350 r_{\rm g}$.

We estimate the mass accretion rate $\dot{M}$ based on the MCD
parameters. Substituting $T_\mathrm{{in}}\sim 5.9 \times 10^{-4}$ keV and
$R_\mathrm{{in,SED}}\sim 6.3 \times 10^5$ km into the equation \citep{kat08} 
\begin{linenomath}$$
\sigma T_\mathrm{in}^4 \simeq \frac{3}{8\pi} \frac{GM\dot{M}}{R_\mathrm{in}^3},
$$\end{linenomath}
we obtain 
\begin{linenomath}$$
\dot{m} \sim 0.04 (M/7 M_\odot)^{-2}.
$$\end{linenomath}
However, we find that this mass accretion rate is too large to explain the
observed optical and X-ray luminosities; 
assuming that the inner accretion disk is RIAF, 
its optical and X-ray luminosities are predicted to be $3 \times 10^{36}$ ergs s$^{-1}$ and
$2 \times 10^{35}$ ergs s$^{-1}$, respectively,
by using the \citet{man97} SED and the scaling laws by \citet{mah97}
(see Section~\ref{sec:discussion1}). Thus, we conclude that the truncated hot standard-disk
scenario is unlikely.

\subsection{Origin of the Optical-UV Power-Law Component}
\label{sec:discussion3}

In addition to the thermal synchro-cyclotron emission from the RIAF and the
blackbody from the companion star, a blue power-law component with
$\Gamma = 1.4^{+0.4}_{-0.9}$ dominating the UV flux is required from the SED
analysis. The slope is consistent with that in the optical band
observed in the initial outburst ($\Gamma=1.7$), which was interpreted
to be jet emission \citep{sdt18}. We thus infer that
this optical-UV power-law component is originated from non-thermal
synchrotron emission from the jet. Adachi et al.\ (in prep.) detected
rapid variability in the optical band in the same re-brightening
phase. The energy spectrum of the variability can be represented with
a power law whose slope ($\Gamma \approx 1.2-1.5$) is consistent with
ours. The presence of the power-law component in the optical SED is
consistent with their results. Generally, compact jets are common at
low mass accretion rates, when the innermost accretion flow becomes
RIAF, including ``hot flow'' in the LHS.

Although the X-ray spectrum is also well fit with a power-law model
with a similar slope, the two power components observed in the
optical-UV and X-ray bands do not smoothly connect each other. This
indicates that they do not originate from a single component (i.e.,
synchrotron radiation from the jet). This is in line with our
interpretation that the X-ray component mainly originates from the RIAF
itself (Section~\ref{subsubsec:ana_sed}). Nevertheless, we do not rule out the possibility
that the same non-thermal electrons in the jets produce X-rays via
Comptonization of the synchrotron photons, which may partially
contribute to the total X-ray flux.

\subsection{Origin of H$_\alpha$ Emission Line}

We have detected a weak and narrow H$\alpha$ emission line. Similar
features are also observed in the quiescence state of dwarf novae
(e.g., GW Lib, \citet{hir09}). The line width of H$\alpha$ emission
constrains the location of its emitting region. 
We estimate the inner radius to be $\sim 10^5 r_{\rm g}$ with 
the double Gaussian model, which is consistent with the lower limit
obtained with the {\tt diskline} model ($\gtrsim 2\times10^4 r_{\rm g}$). 
If the
accretion disk were also RIAF at these radii, its temperature would be
$\sim10^{7}$ K (e.g., \cite{nar95}), which is too high for H$\alpha$
emission to be observed. Hence, we suggest that the entire disk
structure cannot be described by a single RIAF solution but
cooler material must exist at the outermost region. The H$\alpha$ emission may
be originated from an outer disk in the cool state
via irradiation by the inner RIAF and jet, or from
optically-thin plasma via collisional excitation as discussed in \citet{hir09}. 
The
evolution of the $H\alpha$ profile over the entire outburst epoch will
tell us about its origins, which is left as a future work.

\section{Summary}

We have studied the properties of the accretion disk in the Galactic
X-ray black hole binary MAXI J1820 in its first re-brightening
phase, on the basis of our quasi-simultaneous near-IR, optical, UV and
X-ray observations performed in 2019 May. What we have found can be
summarized as follows.

\begin{enumerate}

\item The optical-to-X-ray luminosity ratio ($\sim 8$) is found to much larger
    than those observed in an earlier outburst epoch when the source was
    in the LHS or HSS\citep{sdt18, sdt19}, suggesting
    a very different disk structure.

\item The primary components of the optical and X-ray SED can be best
    interpreted by a RIAF spectrum, showing a sharp luminosity peak in
    the optical band (via thermal synchro-cyclotron) and a power-law in the
    X-ray band (via thermal Bremsstrahlung). By comparison with
    theoretical calculations, we estimate the mass accretion rate to be
    $\dot{m} (\equiv \dot{M} c^2/8 L_{\rm Edd}) =10^{-3}$, which reproduces
    the observed optical and X-ray luminosities within factors of 6.

\item The truncated hot standard-disk scenario to explain the optical SED is unlikely, 
because the inferred mass accretion rate would largely overproduce the observed luminosities. 

\item In addition to the RIAF emission, a blue power-law component
    dominating the UV flux is detected. We interpret that this is
    produced by synchrotron radiation by non-thermal electrons in the
    jet, as suggested by its rapid variability (Adachi et al.\ in
    prep.).

    \item The optical spectrum observed with KOOLS-IFU on the Seimei
      telescope shows a weak H$\alpha$ emission line with an FWHM of
      $\sim 5.3 \times 10^2$ km s$^{-1}$.  The inner radius of the
      H$\alpha$ emitting region is constrained to be $\gtrsim 2
      \times 10^{4} r_{\rm g}$. We suggest that the entire disk
      structure cannot be described by a single RIAF solution but
      cooler material must exist at the outermost region.

\end{enumerate}

\begin{ack}
We thank Kazuya Matsubayashi for his help in the Seimei data reduction.
We acknowledge the use of {\maxi} data provided by RIKEN, JAXA 
and the {\maxi} team, and of public data from the {\swift} 
data archive. Part of this work was financially supported 
by Grants-in-Aid for Scientific Research 19K14762 (MS), 
17H06362 (YU, HN, NK)
from the Ministry of Education, Culture, Sports, Science and 
Technology (MEXT) of Japan. 
Optical and Near-Infrared Astronomy Inter-University Cooperation
Program, supported by the MEXT of Japan
\end{ack}

\bibliography{j1820_ty}{}
\bibliographystyle{aasjournal}

\if0

\fi



\end{document}